\documentclass[10pt, conference, compsocconf]{IEEEtran}
\ifCLASSINFOpdf
\else
\fi

\usepackage{fixltx2e}

\usepackage{float}
\newcommand{\eg}{{\it e.g., }}

\newcommand{\ie}{{\it i.e., }}
\newcommand{\name}{{ClustCrypt}}

\newcommand{\comments}[1]{}
\newcommand\hl{\bgroup\markoverwith
  {\textcolor{yellow}{\rule[-.5ex]{2pt}{2.5ex}}}\ULon}
\usepackage{ amsmath, amssymb, url, lscape, subfigure, algorithmic, multirow, pslatex, listings, verbatim, alltt, amsfonts, wrapfig, boxedminipage, color, cite}
\usepackage[vlined,linesnumbered,ruled,boxed]{algorithm2e}
\usepackage{balance}
\usepackage[dvips]{graphicx}
\usepackage{epsfig}

\usepackage{xcolor}
\usepackage{leqno}
\usepackage[labelformat=parens,labelsep=quad, skip=3pt]{caption}
\usepackage [autostyle, english = american]{csquotes}
\usepackage{multicol}
\usepackage[export]{adjustbox}
\begin{document}
%
\title{ClustCrypt: Privacy-Preserving Clustering of Unstructured Big Data in the Cloud}



\author{\IEEEauthorblockN{SM Zobaed\IEEEauthorrefmark{1},
Sahan Ahmad\IEEEauthorrefmark{1},
Raju Gottumukkala\IEEEauthorrefmark{2}, and
Mohsen Amini Salehi\IEEEauthorrefmark{1}}

\IEEEauthorblockA{\IEEEauthorrefmark{1}School of Computing \& Informatics 
}
\IEEEauthorblockA{\IEEEauthorrefmark{2}Informatics Research Institute\\
University of Louisiana at Lafayette, Louisiana 70504, USA\\Email: \{sm.zobaed1, sahan.ahmad1, raju, amini\}@louisiana.edu
}
}

\maketitle
\vspace{-5pt}

\begin{abstract}
Security and confidentiality of big data stored in the cloud are important concerns for many organizations to adopt cloud services. One common approach to address the concerns
 is client-side encryption where data is encrypted on the client machine before being stored in the cloud. Having encrypted data in the cloud, however, limits the ability of data clustering, which is a crucial part of many data analytics applications, such as search systems. 
To overcome the limitation, in this paper, we present an approach named \name~for efficient topic-based clustering of encrypted unstructured big data in the cloud.
\name~dynamically estimates the optimal number of clusters based on the statistical characteristics of encrypted data. It also provides clustering approach for encrypted data. 
We deploy \name~within the context of a secure cloud-based semantic search system (S3BD). Experimental results obtained from evaluating \name~on three datasets demonstrate on average 60\% improvement on clusters' coherency. 
  \name~also decreases the search-time overhead by up to 78\% and increases the accuracy of search results by up to 35\%.
 

\end{abstract}


%
\IEEEpeerreviewmaketitle

\section{Introduction}\label{sec:intro}

Many organizations own high volume of unstructured documents in forms of reports, emails, or web pages. The size of these data is expected to reach 44 zetabytes within the next two years~\cite{infoot}. Cloud providers offer scalable and convenient services to store, process, and analyze the massive volume of data, which is also known as big data. As such, organizations use cloud services to relieve from the burden of storing large volume of data locally. However, the outsourced contents (documents) often contain private or sensitive information (\eg criminal or financial reports~\cite{zobaedbig}) that need to be protected against internal and external cloud attackers. In fact, data security and privacy concerns have made many organizations reluctant to use cloud. For instance, in 2018, more than 14 million Verizon customers' accounts information were leaked from their cloud repository~\cite{verizonacci}.                                                                 
In another instance, confidential information of more than three billion Yahoo users were exposed~\cite{yahooacci}. In~\cite{cloudacci2}, numerous similar data privacy violation incidents are reported. These incidents vividly highlight the importance of this issue in the cloud era.
 
An ideal privacy preserving cloud solution is expected to enable organizations to securely utilize cloud services, while providing access and search ability only to authorized users. Such a solution should be lightweight and users can have it on their thin-clients (\eg smartphones) with storage, energy, and data processing constraints. Client-side encryption~\cite{S3C}, in which documents are encrypted with private keys before outsourcing them to the cloud, is a promising approach to achieve data privacy. However, the ability to process and search the encrypted documents is lost.

Searchable Encryption systems (\eg~\cite{pham2018survey,cao2014privacy}) have been developed to enable privacy preserving search ability over encrypted data. Such systems predominantly extract keywords (aka tokens) from documents and leverage them to carry out the search operation. The extracted keywords and the documents they appeared in are mapped to an index structure~\cite{S3C, cao2014privacy}, which is then traversed against a search query to find relevant documents. A problem arises for big datasets where the index becomes the processing bottleneck to conduct real-time search operation~\cite{S3BD}. For instance, in S3C~\cite{S3C} search tool, for a 100-petabyte dataset the index size grows to $\approx$ 300 GB. Traversing such a large index affects searching timeliness 
and can be potentially impacted by hardware limitations~\cite{salehi2017reseed}. To resolve the bottleneck, the index structure can be partitioned to several \emph{clusters}~\cite{XuCroft, Google}. Then, for a given search query, the search space is pruned and limited to the clusters that are relevant to the query. 

Although numerous data clustering methods exist, they are not appropriate for encrypted big data because of the following challenges: \emph{First,} in the encrypted domain the original data is not available. Therefore, prior works (\eg \cite{XuCroft, Google, nocs2}) suggest making use of statistical characteristics as the clustering metric for encrypted data. For instance, in S3BD~\cite{S3BD}, which is a search system for encrypted big data, keywords' co-occurrences in a document set is used to cluster keywords. However, in S3BD, a constant number of clusters ($k$) is considered, regardless of the dataset characteristics. \emph{Second,} complexity of the clustering methods (\eg \textit{K}-means~\cite{kmeans}) make them unscalable for big data~\cite{aggarwal2001surprising}.

To address these two challenges, in this research, we investigate \emph{how to optimally and scalably cluster keywords in encrypted big datasets?}. The outcome of this research is \name~that enhances clustering of encrypted keywords by estimating the appropriate number of clusters ($k$) and distributing keywords across them. Unlike traditional \textit{k}-means-based clustering approaches, where ($k$) is chosen arbitrarily \cite{kmeans}, \name~estimates $k$ by probabilistically inferring the underlying semantic similarity among the encrypted tokens. The probabilistic calculation measures the tendency for each token to be separate from others. It is noteworthy that \name~has the advantage of improving the search quality and timeliness without implying any architectural changes to the existing systems. We develop \name~and evaluate it on three different datasets. We compare and analyze the the number and coherency of resulting clusters against baseline and state-of-the-art clustering approaches that operate on plain-text data.

In summary, contributions of this paper are as follows:
\begin{itemize}
 \item Estimating the appropriate number of clusters ($k$) for a given encrypted dataset based on its characteristics.
 \item Proposing a method to distribute encrypted keywords to relevant clusters.
 \item Evaluating clusters in terms of coherency and their impact on the search results. 
\end{itemize}

The rest of the paper is organized as follows. In Sections \ref{sec-background} and \ref{sec:back}, we discuss related works and background. We explain \name~in Section \ref{sec: propo}. Then, in Section \ref{sec: seca}, we discuss the threat model and provide security analysis of \name. Sections \ref{sec:evltn} and \ref{sec:conclsn} explain results and conclusion.

\section{Related Works}
\label{sec-background}
 
Multiple methods have been provided to estimate the suitable number of clusters ($k$), and clustering processes. Determining the optimum number of clusters \textit{k} is an NP-hard problem~\cite{Paparrizos}.
Hence, 
research has been undertaken to provide heuristic methods in which $k$ is approximated and clustering approaches. 
In this section, we review such prior studies and position our work corresponding to them. 

Many clustering approaches are independent from the dataset they are applied on. Such generic clustering methods, such as \textit{k}-means, populate a predefined set of $k$ clusters based on the convergence of center shifting~\cite{lee2009pca},~\cite{pelleg2000x}. For a dataset with $n$ data points, $k$ clusters, $I$ iterations, and $m$ attributes, the time complexity of \textit{k}-means is $O(n \times k\times I\times m)$, which is not scalable for big data \cite{vrahatis2002new, kcomplexity}. 
In contrast, after determining $k$, \name~populates clusters by achieving the two following steps. First, from the set of data points, we find appropriate centers for each cluster and assign each token to proper cluster. 
 The overall time complexity of the two steps is $O(n^2)$.

Pelleg and Moore~\cite{pelleg2000x} proposed a framework, named \textit{x}-means clustering, to approximate $k$ via learning methods. The framework is a modified version of \textit{k}-means clustering that starts with one cluster (\ie \textit{k}=1) and iteratively subdivides it maintaining stopping criterion~\cite{comjnl/32.3.193}. An optimization function is used to produce the appropriate $k$ to capture the variety exists in the data points. 
At some value of \textit{k}, when the function reaches a plateau, it is assumed that the starting point of the plateau is the optimal $k$~\cite{coates2012learning}. However, in this framework, it is difficult to apply the stopping criterion~\cite{comjnl/32.3.193}, particularly, if no prior knowledge on dataset is available.

Progeny~\cite{hu2015progeny} is a clustering method that constructs new clusters out of existing ones based on their proposed stability metric. The measure of stability in this method is based on a co-occurrence probability matrix that verifies the appropriate cluster for new data points. The method starts with a range of possible solutions (\ie set of \textit{k} values). The optimal value of \textit{k} is determined based on the highest stability score. We compare \name~against with this method in Section \ref{sec:evltn}.


Methods utilizing genetic algorithm~\cite{lleti2004selecting} have shown to perform well in solving different optimization problems, such as clustering, by starting with a population of a set of potential solutions (chromosomes) and evolving towards a nearly optimal solution. 
Instead of providing several solutions to a particular problem, genetic algorithms keep a well balance between exploration (crossover) of the space of solutions and the exploitation (mutation) of the promising regions. 
However, it is computationally prohibitive to utilize Genetic algorithms for clustering, as we need to consider trillions of combinations generated by tokens of a big dataset. 

The idea of topic-based term clustering on an index structure (\ie data points) has been extensively studied on plain-text. Xu and Croft \cite{XuCroft} proposed the idea that clusters formed for a homogeneous index (\ie an index that all of its terms share a similar topic) improve the efficacy of a search system in comparison to standard state-of-the-art information retrieval systems. The authors used \textit{k}-means clustering method and distributed the indexed elements among the clusters using KL-divergence distance function \cite{kullback1951information}. Once the clusters are built, for an incoming search query, they perform pruning to determine the highest relevant cluster via utilizing the maximum likelihood estimation theory.


\section{Background}\label{sec:back}
\name~is motivated by S3BD, a cloud-based secure semantic search system, that requires clustering over encrypted big data~\cite{S3BD}. The search system can be used by law-enforcement agencies to detect criminal activities by  searching over privacy-protected criminal reports.

\begin{figure}
	\centering
    
	\includegraphics[width=.92\linewidth]{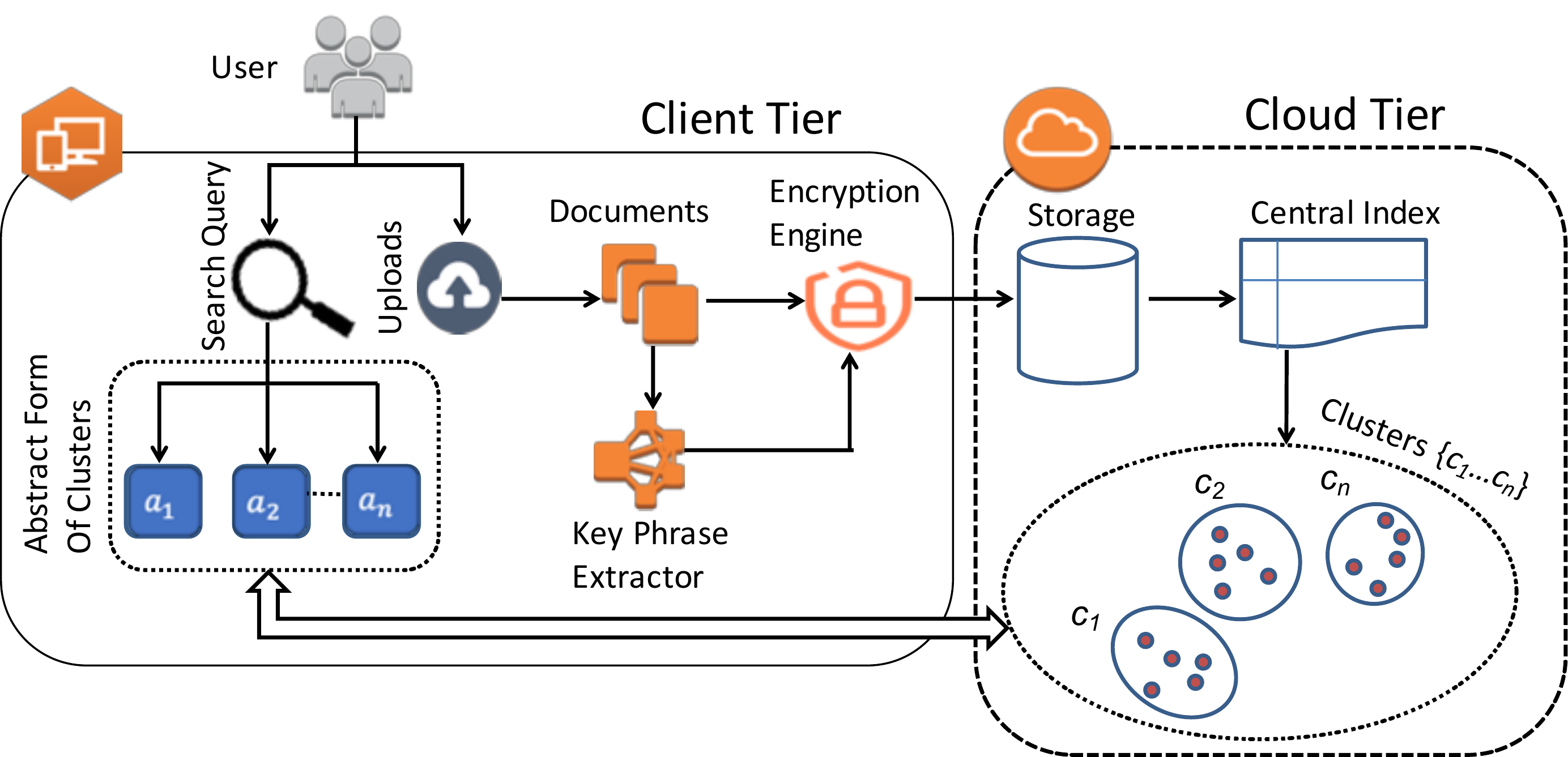}
    
	\caption{Overall Search System Architecture Integrating Proposed Clustering Approach. }
	\label{sd}
\end{figure}

Figure~\ref{sd} presents where \name~is positioned within the S3BD system. We can see that S3BD is composed of a \emph{client tier} and a \emph{cloud tier}. The client tier is considered trusted and it provides \emph{upload} and \emph{search} functionalities for the users. The cloud tier is considered honest but curious, therefore, all the documents and their indexed tokens are stored in encrypted form. To enable real-time searching, the encrypted indexed tokens have to be clustered. 

In the current version, S3BD repurposes \textit{k}-means clustering to restrict the search space and provide real-time search operation over big data. However, the clustering is carried out based on a fixed \textit{k} value, regardless of the dataset characteristics that limits its search quality. In this research, we propose an approach that considers the dataset characteristics and improves clustering of the encrypted keywords. 
Although we implemented \name~clustering approach in the context of S3BD, the approach is generic and can be deployed in other systems that require clustering of encrypted data. 

As shown in Figure~\ref{sd}, once a user uploads documents, a keyword extractor is used to extract \textit{n} keywords (aka tokens) from the original documents. S3BD fairly treats all documents and keeps the value of \textit{n} constant across all documents in a dataset. Then, the documents and tokens are both encrypted and sent to the cloud tier. RSA deterministic encryption technique \cite{S3BD} is used to encrypt documents and extracted tokens. 

Cloud tier maintains a \emph{central index} structure that includes key-value pairs. Each key-value pair represents, respectively, an encrypted token, and the list of documents where the token appears in, plus the frequency of the token in each one of those documents. Homomorphic encryption~\cite{homomorphic:slow} can be used to encrypt the token frequency information. However, due to the slow down imposed by processing homomorphically encrypted data~\cite{homomorphic:slow} and to practically maintain the real-time search quality, currently, the frequency information are stored in unencrypted form. Upon issuing a search query by a user, the search keywords are encrypted and searched against the central index in the cloud tier to retrieve relevant documents. Upon receiving the list of matching documents, the user can download and decrypt them utilizing his/her private key.

Clusters $c_1,...,c_n$ are constructed based on the tokens of the index structure and to mitigate exhaustively searching the whole index structure for every single search query. The clusters are topic-based and they are constructed so that the union of the \textit{k} clusters is equivalent to the index structure. For a given search query, instead of searching the whole index, the search space is pruned and gets limited to only those clusters that are topically related to the search query. The pruning is achieved based on a set of Abstract structures (denoted $a_1,...,a_n$) that are sampled from each one of the clusters and reside either on the client tier or possibly on a trusted edge server~\cite{salehi2017reseed}. 
Details of the way abstracts are created described in~\cite{S3BD}~\cite{usercentri2019}. Upon issuing a search query, the most similar abstracts to the search query are chosen and then, their corresponding clusters are searched. 


\section{\name: Clustering Encrypted Tokens}
\label{sec: propo}
\subsection{Overview}
Topic-based clustering partitions indexed tokens based on their similarity. Due to encryption, however, the indexed tokens do not carry any semantic information that makes clustering a challenging task. To overcome this challenge, clustering is achieved based on \emph{statistical semantics}. The idea is to locate tokens that are semantically close to each other in the same cluster. To achieve this, we first need to know the number of clusters ($k$) that should be created to cover topics exist in token of a given dataset. Then, we find the central tokens for each cluster and assign the rest of tokens to the most topically related clusters. 
In our work, according to S3BD system, we consider that the frequency and co-occurrences of all tokens in the dataset are available. 

In the rest of this section, we first describe how to estimate the appropriate ($k$) for a given set of indexed tokens. Then, in Sections \ref{3.3} and \ref{cluster-wise}, we explain the center selection  and token distribution methods. 

\subsection{Estimating Number of Clusters ($k$)}
\label{3.2}
Determining the appropriate number of clusters ($k$) is important, because it directly impacts the searching performance. Depending on the characteristics of a dataset and distribution of tokens in its documents, the value of $k$ can vary significantly. Encrypted tokens, extracted from documents, and statistical data corresponding to them (\ie tokens' appearance in documents and their frequency) are available parameters to estimate $k$ in the following manner:

\noindent \textbf{Step-1: Building Token-Document Frequency Matrix.}
We initialize a token-document matrix $A$ from the index structure. In the matrix, each row represents a token and each column represents a document. To be able to follow the method throughout the paper, we consider an example using \textit{five} tokens and \textit{six} documents in Table \ref{ini mat}. Although our approach does not deal with plain-text tokens, for further readability, in this table, we redundantly show plain-text tokens along with the encrypted ones.
Each entry $a_{i,j}$ of matrix $A$ represents the frequency of $i^{th}$ token in $j^{th}$ document (denoted as $f(i,j)$). 

\begin{table}[h]
\centering
\caption{Token-Document Matrix $A$ obtained from index}
\label{ini mat}
\begin{tabular}{|l|l|c|c|c|c|c|c|}
\hline
\textbf{Word}       & \textbf{Hash}                                      & \multicolumn{1}{l|}{\textbf{d\textsubscript{1}}} & \multicolumn{1}{l|}{\textbf{d\textsubscript{2}}} & \multicolumn{1}{l|}{\textbf{d\textsubscript{3}}} & \multicolumn{1}{l|}{\textbf{d\textsubscript{4}}} & \multicolumn{1}{l|}{\textbf{d\textsubscript{5}}} & \multicolumn{1}{l|}{\textbf{d\textsubscript{6}}} \\ \hline

\textcolor {gray}{Book}  & \textbf{Uh5W}                                   & 30                               & 0                                & 23                               & 4                                & 40                               & 0                                \\ \hline
\textcolor {gray}{Solve} & \textbf{/Vdn}                                    & 5                                & 0                                & 0                                & 60                               & 34                               & 0                                \\ \hline
\textcolor {gray}{Traffic}    & \textbf{oR1r}                                  & 0                                & 23                               & 0                                & 30                               & 0                                & 0                                \\ \hline
\textcolor {gray}{Net}    & \textbf{vJHZ} & 52                               & 49                               & 0                                & 23                               & 0                                & 26                               \\ \hline
\textcolor {gray}{Enter}    & \textbf{tH7c}                                  & 0                                & 45                               & 68                               & 0                                & 3                                & 5                                \\ \hline

\end{tabular}
\end{table}

For a big dataset, the matrix size can be prohibitively large and sparse. To avoid this, we trim it to include only the token that are influential in building clusters. We define \textit{document co-occurrences} as the number of documents containing a particular token. Then, to build the token-document frequency matrix ($A$), we only consider tokens whose document co-occurrences are either greater than or equal to the mean value of the document co-occurrences across the whole dataset. 

\noindent \textbf{Step-2: Constructing Normalized Matrix.}
To make the relationship among tokens and documents quantifiable and comparable, we need to normalize the token-document frequency matrix. Considering that $a_{i,j}$ represents the strength of association between token $i$ and document $j$, the maximum value in column $j$ of the token-document frequency matrix represents the token with the highest association with $j$. 

Therefore, for normalization, we divide the value of each entry of matrix $A$ to the highest value in the corresponding column of the matrix and the result is stored in a matrix, called $N$. The value for each entry $n_{i,j}$ is formally calculated by $\frac{a_{i,j}}{\displaystyle\max_{\forall i} a_{i,j}} $.  
%
For the example provided in Table~\ref{ini mat}, the normalized matrix $N$ is represented in Table \ref{Normalized_mat}.

\begin{table}[H]

\centering
\vspace{-1pt}
\caption{Normalized Token-Document matrix $N$}
\label{Normalized_mat}
\begin{tabular}{|l|l|c|c|c|c|c|c|}
\hline
\textbf{Word}       & \textbf{Hash}                                      & \multicolumn{1}{l|}{\textbf{d\textsubscript{1}}} & \multicolumn{1}{l|}{\textbf{d\textsubscript{2}}} & \multicolumn{1}{l|}{\textbf{d\textsubscript{3}}} & \multicolumn{1}{l|}{\textbf{d\textsubscript{4}}} & \multicolumn{1}{l|}{\textbf{d\textsubscript{5}}} & \multicolumn{1}{l|}{\textbf{d\textsubscript{6}}} \\ \hline
\textcolor {gray} {\scriptsize{Book}}  & \textbf{\scriptsize{Uh5W}}                                   & 0.58                              & 0                                & 0.34                              & 0.07                              & 1                                & 0                                \\ \hline
\textcolor {gray}{\scriptsize{Solve}} & \textbf{\scriptsize{/Vdn}}                                    & 0.1                               & 0                                & 0                                & 1                                & 0.85                              & 0                                \\ \hline
\textcolor {gray}{\scriptsize{Traffic}}    & \textbf{\scriptsize{oR1r}}                                  & 0                                & 0.47                              & 0                                & 0.5                               & 0                                & 0                                \\ \hline
\textcolor {gray}{\scriptsize{Net}}    & \textbf{\scriptsize{vJHZ}} & 1                                & 1                                & 0                                & 0.38                              & 0                                & 1                                \\ \hline
\textcolor {gray}{\scriptsize{Enter}}    & \textbf{\scriptsize{tH7c}}                                  & 0                                & 0.92                              & 1                                & 0                                & 0.08                              & 0.19                              \\ \hline
\end{tabular}

\end{table}

\noindent \textbf{Step-3: Building Probabilistic Matrices $R$ and $S$. }
The goal, in this step, is to calculate the topic similarity among encrypted tokens. 
For that purpose, we need to calculate the probability that topic of a token shares similarity with other tokens. Our hypothesis is that tokens that co-occur across documents are likely to share the same topic. In addition, the magnitude of similarity between two given tokens could be influenced by distribution of the tokens across the dataset. For instance, specific terms 
appear only in a few documents and are not widely distributed throughout the dataset. 
Such sparsely distributed tokens have low co-occurrences with other tokens which increases the diversity of topics in a dataset and potentially raises the number of required clusters ($k$). 
We leverage the normalized matrix ($N$) to perform a two-phase probability calculation that ultimately yields a matrix (denoted as $C$) representing token-to-token topic similarity. 

In the first phase, we calculate the \emph{importance} of each token to each documents. The importance of token $t_i$, in document $d_j$ denoted by $\tau_{ij}$ and is defined as $\tau_{ij}$ $=$ ${n_{i,j}}/{\displaystyle\sum_{\forall k} n_{i,k}}$. 
 Considering each $\tau_{ij}$ 
 and N, we generate $R$ matrix whose entries represent the importance of each token across all documents. In fact, each entry $r_{i,j}$ of $R$ matrix represents the probability of choosing a document $d_j$, having token $t_i$. That is, $r_{i,j}=P(t_i,d_j)$.

 In our example, Table \ref{R mat} shows the $R$ matrix obtained from the $N$ matrix (shown in Table~\ref{Normalized_mat}). 


\begin{table}[H]
\centering
\vspace{-2pt}
\caption{$R$ matrix: Built from N} 
\label{R mat}
\begin{tabular}{|l|l|c|c|c|c|c|c|}
\hline
\textcolor {gray}{\scriptsize{Word}}       & \textbf{\scriptsize{Hash} }                                     & \multicolumn{1}{l|}{  \textbf{\scriptsize{d\textsubscript{1}}}} & \multicolumn{1}{l|}{\textbf{ \scriptsize{ d\textsubscript{2}}}} & \multicolumn{1}{l|}{\textbf{  d\textsubscript{3}}} & \multicolumn{1}{l|}{\textbf{\scriptsize{  d\textsubscript{4}}}} & \multicolumn{1}{l|}{\textbf{ \scriptsize{ d\textsubscript{5}}}} & \multicolumn{1}{l|}{  \textbf{\scriptsize{  d\textsubscript{6}}}} \\ \hline
\textcolor {gray}{\scriptsize{Book}}  & \textbf{\scriptsize{Uh5W}}                                   & 0.29                              & 0                                & 0.17                              & 0.04                              & 0.50                              & 0                                \\ \hline
\textcolor {gray}{\scriptsize{Solve}} & \textbf{\scriptsize{/Vdn}}                                    & { .05}                        & { 0}                          & { 0}                          & { 0.51}                        & {0.43}                        & { 0}                          \\ \hline
\textcolor {gray}{\scriptsize{Traffic}}    & \textbf{\scriptsize{oRir} }                                 & { 0}                          & { 0.48}                        & { 0}                          & { 0.52}                        & { 0}                          & { 0}                          \\ \hline
\textcolor {gray}{\scriptsize{Net}}    & \textbf{\scriptsize{vJHZ}} & {0.29}                        & {0.29}                        & {0}                          & { 0.11}                          & { 0}                        & {0.29}                        \\ \hline
\textcolor {gray}{\scriptsize{Enter} }   & \textbf{\scriptsize{tH7c}}                                  & { 0}                          & { 0.42}                        & { 0.45}                        & { 0}                          & { 0.03}                        & { 0.09}                        \\ \hline
\end{tabular}
\end{table}

In the second phase, we calculate the importance of each document to each token. The importance of document $d_j$ for term $t_i$, denoted by $\delta_{ji}$ and is defined as $\delta_{ji}$ $=$ ${n_{j,i}}/{\displaystyle\sum_{\forall q} n_{q,i}}$. 
Considering each $\delta_{ji}$ 
and N, we generate $S$ matrix whose entries represent the importance of each document with respect to each token. In fact, each entry $s_{i,j}$ represents the probability of choosing $t_i$ from $d_j$ (\ie we have $s_{i,j}=P(d_j,t_i)$). 
In our example, Table \ref{s mat} shows the $S$ matrix obtained from the $N$ matrix (shown in Table~\ref{Normalized_mat}). 

\begin{table}[H]
\vspace{-5 pt}
\centering
\caption{$S$ matrix: Built from N} 
\label{s mat}
\begin{tabular}{|l|c|c|c|c|c|}
\hline
 \textbf{Docs} & \multicolumn{1}{l|}{\textbf{\begin{tabular}[c]{@{}l@{}}\textcolor {gray}{Book}\\
 Uh5W\end{tabular}}} & \multicolumn{1}{l|}{\textbf{\begin{tabular}[c]{@{}l@{}}\textcolor {gray}{Solve}\\
 /Vdn\end{tabular}}} & \multicolumn{1}{l|}{\textbf{\begin{tabular}[c]{@{}l@{}} \textcolor {gray}{Traffic}\\
  oRir\end{tabular}}} & \multicolumn{1}{l|}{\textbf{\begin{tabular}[c]{@{}l@{}} \textcolor {gray}{Net}\\
 vJHZ\end{tabular}}} & \multicolumn{1}{l|}{\textbf{\begin{tabular}[c]{@{}l@{}} \textcolor {gray}{Enter}\\
 tH7c\end{tabular}}} \\ \hline
\textbf{d\textsubscript{1}} & 0.34                                                                                       & 0.06                                                                                       & 0                                                                                      & 0.60                                                                                                                       & 0                                                                                        \\ \hline
\textbf{d\textsubscript{2}} & 0                                                                                         & 0                                                                                         & 0.19                                                                                    & 0.49                                                                                                                       & 0.38                                                                                      \\ \hline
\textbf{d\textsubscript{3}} & 0.17                                                                                       & 0                                                                                         & 0                                                                                      & 0                                                                                                                         & 0.45                                                                                      \\ \hline
\textbf{d\textsubscript{4}} & .04                                                                                       & 0.51                                                                                       & 0.25                                                                                    & 0.19                                                                                                                       & 0                                                                                        \\ \hline
\textbf{d\textsubscript{5}} & 0.52                                                                                       & 0.44                                                                                       & 0                                                                                      & 0                                                                                                                         & 0.04                                                                                      \\ \hline
\textbf{d\textsubscript{6}} & 0                                                                                         & 0                                                                                         & 0                                                                                      & 0.84                                                                                                                       & 0.16                                                                                      \\ \hline
\end{tabular}
\end{table}

\noindent \textbf{Step-4: Determining Number of Clusters.}
Recall that $R$ is a token-to-document matrix and $S$ is a document-to-token matrix. To identify the similarity among the encrypted tokens, we multiply $R$ and $S$ matrices. As the number of columns and rows of $R$ and $S$ are equal, it is possible to multiply matrix $R$ with $S$. The resultant matrix of multiplying $R$ and $S$, denoted as $C$, is a token-to-token matrix and serves as the base to determine the number of required clusters. Each entry $c_{i,j}$ denotes the topic similarity between token $i$ and $j$. More specifically, $c_{i,j}$ that indicates the magnitude to which token $i$ shares similar topic with token $j$ for $i\neq j$ is calculated as $c_{i,j}= \displaystyle\sum_{\forall i,j}r_{i,j}\cdotp s_{j,i}$. 
 Table \ref{c mat} shows the example of Matrix $C$ which is used throughout this section.



\begin{table}[H]
\centering

\caption{C matrix: Multiplication of R \& S }
\label{c mat}
\begin{tabular}{|l|c|c|c|c|c|}

\hline
\textbf{Word - Hash}                                                                                       & \multicolumn{1}{l|}{\textbf{\begin{tabular}[c]{@{}l@{}}\textcolor {gray}{Book}\\ Uh5W\end{tabular}}} & \multicolumn{1}{l|}{\textbf{\begin{tabular}[c]{@{}l@{}} \textcolor {gray}{Solve}\\ /Vdn\end{tabular}}} & \multicolumn{1}{l|}{\textbf{\begin{tabular}[c]{@{}l@{}} \textcolor {gray}{Traffic}\\ oRir\end{tabular}}} & \multicolumn{1}{l|}{\textbf{\begin{tabular}[c]{@{}l@{}} \textcolor {gray}{Net}\\ vJHZ\end{tabular}}} & \multicolumn{1}{l|}{\textbf{\begin{tabular}[c]{@{}l@{}} \textcolor {gray}{Enter}\\  tH7c\end{tabular}}} \\ \hline
\textbf{\begin{tabular}[c]{@{}l@{}}\scriptsize{\textcolor {gray}{Book}- Uh5W}\end{tabular}}                                       & \textbf{0.39}                                                                                  & 0.25                                                                                  & 0.01                                                                                     & 0.18                                                                                                                       & 0.09                                                                                     \\ \hline
\textbf{\begin{tabular}[c]{@{}l@{}}\scriptsize{\textcolor {gray}{Solve}- /Vdn}\end{tabular}}                                      & 0.26                                                                                  & \textbf{0.45}                                                                                  & 0.12                                                                                     & 0.12                                                                                                                       & 0.02                                                                                     \\ \hline
\textbf{\begin{tabular}[c]{@{}l@{}}\scriptsize{\textcolor {gray}{Traffic}-\ oRir}\end{tabular}}                                  & 0.02                                                                                  & 0.26                                                                                  & \textbf{0.21}                                                                                     & 0.33                                                                                                                       & 0.18                                                                                     \\ \hline
\textbf{\begin{tabular}[c]{@{}l@{}}\scriptsize{\textcolor {gray}{Net}- vJHZ}\end{tabular}} & 0.10                                                                                  & 0.07                                                                                  & 0.08                                                                                     & \textbf{0.58}                                                                                                                       & 0.15                                                                                     \\ \hline
\textbf{\begin{tabular}[c]{@{}l@{}}\scriptsize{\textcolor {gray}{Enter}-  tH7c}\end{tabular}}                                  & 0.09                                                                                  & 0.01                                                                                  & 0.08                                                                                     & 0.28                                                                                                                       & \textbf{0.37}                                                                                     \\ \hline
\end{tabular}
\end{table}
\vspace{-5pt}

Diagonal entries of $C$ signify the topic similarity of each tokens with itself and separation from other topics. More specifically, the value of $c_{i,i}$ indicates the magnitude that term $t_i$ does not share its topic with other terms. Therefore, we define diagonal entries ($c_{i,i}$) as \textit{separation factor}, because for each token it represents the token's tendency to stay separate from other topics to the degree of the coefficient. As such, summation of the separation factors can approximate the number of clusters $k$ needed to partition topics of a dataset. Let $m$ denote the total number of tokens in matrix $C$. Then, Equation~\ref{eq:trace} is used to approximate $k$ for a dataset. We use the ceiling function to make $k$ an integer value.

\begin{equation}
    k = \lceil\sum_{i=1}^{m} c_{i,i}\rceil
\label{eq:trace}
\end{equation}

Correctness of $k$ is verified using a hypothesis that states $k$ within a set should be higher if individual elements of the set are dissimilar, otherwise $k$ should be lower~\cite{Can1990,cutting2017scatter}. Equation~\ref{eq:trace} is the core part of approximating $k$. According to this equation, the highest $k$ could be $m$ when each individual token represents a topic, otherwise it is lower. Hence, our approach conforms with the clustering hypothesis.

\subsection{Determining Clusters' Centers}\label{3.3}
In \textit{k}-means clustering, generally, the clusters' centers are arbitrarily chosen~\cite{aggarwal2019performance, LiuCroft}. Then, based on a distance measure function (\eg Euclidean distance~\cite{aggarwal2019performance} or semantic graph \cite{LiuCroft}), dataset elements are distributed into clusters. \textit{K}-means operates based on iteratively shifting clusters' centers until convergence. However, we realized that the extremely large number of tokens make the iterative center shifting step (and therefore \textit{k}-means clustering) prohibitively time consuming for big data~\cite{aggarwal2001surprising}. Accordingly, in this part, we are to propose a big-data-friendly method to cluster encrypted tokens. 
 
The key to our clustering method is to dismiss the iterative center shifting step. This change entails initial clusters' centers not to be chosen arbitrarily, instead, they have to be chosen proactively so that they cover various topics of the dataset. For that purpose, a na\"{i}ve method can be choosing the top \textit{k} tokens from index that have the highest number of associated documents. Although this approach chooses important (highly associated) tokens, it selects centers with document and topical overlap. Alternatively, we propose to choose tokens that not only have high document association, but also cover diverse topics exist in the dataset. 

We define \emph{centrality} of a token $i$, denoted $\Phi_i$, as a measure to represent a topic and relatedness to other tokens of the same topic. Assume that tokens are sorted descendingly based on the degree of document association. Let $U$ represent the union of documents associated to the currently chosen centers. Also, for token $i$, let $A_i$ represent the set of documents associated to $i$. Then, \emph{uniqueness}~\cite{S3BD} of token $i$, denoted $\omega_i$, is defined as the ratio of the number of documents associated to $i$ but not present in $U$ (\ie $|A_i-U|$) to the number of documents associated to $i$ and are present in $U$ (\ie $|A_i\cap U|$). Uniqueness indicates the potential of a token to represent a topic that has not been identified by other tokens already chosen as centers. Particularly, tokens with uniqueness value greater than $1$ have high association to documents that are not covered by the currently chosen centers, hence, can be chosen as new centers.

Recall that each entry $c_{i,j}$ of matrix $C$ represents the topic similarity between tokens $i$ and $j$. Besides, diagonal entry $c_{i,i}$ measures separation of token $i$ from others. Therefore, the total similarity token $i$ shares with others can be obtained by $\Sigma_{\forall j | j \neq i}c_{i,j}$. Note that for token $i$, we have $\Sigma_{\forall j}c_{i,j}=1$, hence, the total similarity for token $i$ is equal to $1-c_{i,i}$. Centrality of a token is measured by the uniqueness of the token, the magnitude of similarity the token shares with others, and the magnitude of it being isolated. That is, for token $i$, centrality is defined as $\Phi_i=\omega_{i}\times c_{i,i}\times (1-c_{i,i})$.  

\begin{algorithm}
\SetAlgoLined\DontPrintSemicolon
\SetKwInOut{Input}{Input}
\SetKwInOut{Output}{Output}
\SetKwFunction{algo}{algo}
\SetKwFunction{proc}{Procedure}{}{}
\SetKwFunction{main}{\textbf{ChooseCenter}}
\SetKwFunction{quant}{\textbf{CalculateUniqueness}}
\Input{$k$, $C$ matrix, and $central$ $index$ (with tokens sorted descendingly based on the degree of document association)}
\Output{Set $centers$ that includes at most $k$ center tokens}

\SetKwBlock{Function}{Function \texttt{ Choose Center($k, C, Index$):}}{end}

\Function{
	$centers \gets \emptyset$ \;
	$U \gets \emptyset$ \; 
	$\Theta \gets \{(\emptyset,\emptyset)\}$ //\small{Pairs of tokens and centrality values} \;
	\ForEach{token $i \in index$} {
		
		$\omega_i \gets \quant(i, U)$\;
     \If {$\omega_i >1$} {	
       	        	        	$U \gets U \cup A_i$ \; 
       	        	        	$\Phi_i \gets (\omega_i \times c_{i,i} \times (1-c_{i,i})) $\;
					Add pair ($i,\Phi_i$) to max-heap $\Theta$ based on $\Phi_i$ \; 
       	        	        }
		
	}

	$centers \gets$ \small{Extract $k$ max pairs from $\Theta$ heap  } \;
	\Return{centers} \;	
	
}
\caption{Determining clusters' centers}
\label{alg:centr}

\end{algorithm}
\vspace{-1pt}
Algorithm \ref{alg:centr} shows the high-level pseudo-code to select maximum of $k$ centers from the set of indexed tokens of a dataset. In addition to $k$, the algorithm receives the central index and the $C$ matrix as inputs. The algorithm returns a set of at most $k$ center tokens, denoted $centers$, as output. We do not consider the lower bound in this case. The algorithm intuitively selects at most K- number of cluster centers. In the beginning, the output set is initialized to null. $U$ represents the set of documents covered with the chosen centers. A max-heap structure, denoted $\Theta$, is used to store a pair for each token and its centrality value. 
For each token $i$, the uniqueness and centrality values are calculated (Steps 5 to 12) and the corresponding pair is inserted to the heap. Note that tokens with uniqueness lower than or equal to one do not have the potential to serve as a cluster center. In the next step, we select at most $k$ center tokens that have the highest centrality values.


%
%
%
%

\subsection{Clustering Tokens}
\label{cluster-wise}

Once $k$ tokens are chosen as cluster centers, the other tokens of the index 
 are distributed among the clusters based on the relatedness (aka distance) between the centers and others. Established techniques exist to calculate such relatedness (\eg semantic graph \cite{LiuCroft}, Euclidean distance~\cite{aggarwal2019performance}), but they are not suitable for tokens that are sparsely distributed~\cite{aggarwal2019performance}.
Besides, these are not designed to apply on encrypted data~\cite{LiuCroft}. 

In S3BD \cite{S3BD}, a method based on document co-occurrence is proposed to measure relatedness and cluster encrypted tokens. In this method, if two tokens are present in the same set of documents, they are considered related~\cite{S3BD}. We utilize that to measure the relatedness of tokens with cluster centers and distribute tokens to the most related cluster. To determine the relatedness between a particular token and a center, we need to calculate the \emph{contribution} and \emph{co-occurrences} metrics for the token. Let $t$ be a token in document $d$ of dataset $D$ with frequency denoted as $f(t,d)$. Then, contribution of $d$ to $t$, denoted as $\kappa(d,t)$, is defined based on Equation~\ref{contribution}. 
 \vspace{-5pt}
 \begin{equation} \label{contribution}
   \vspace{-5pt}
 	\kappa(d, t) = \frac{f(t, d)}{\sum\limits_{j \in D}{f(t,j)}}
 \end{equation}

Co-occurrence of token $t$ with center token $\gamma_x$ in document $d$ (denoted $\rho(t,d,\gamma_x)$) is defined as a ratio of the sum of frequencies of $t$ and center $\gamma_x$ in $d$ to the total frequencies of $t$ and $\gamma_x$ throughout the dataset. The formal presentation of co-occurrence is provided in Equation~\ref{eq:neweq1}. 

 \begin{equation} \label{eq:neweq1}
   \vspace{-5pt}
 \rho(t,d,\gamma_x) = \frac{f(t,d)+f(\gamma_x,d) }
 { \sum\limits_{j \in D}(f(t,j)+f(\gamma_x,j))} 
 \end{equation}

Based on the contribution and co-occurrence metrics, relatedness between token $t$ and $\gamma_x$ (denoted $r(\gamma_x, t)$), is defined as multiplication of these two metrics (shown in Equation~\ref{distance}). 

 \begin{equation} \label{distance}
  \vspace{-5pt}
 	r(\gamma_x, t) = \sum_{j\in D} \kappa (j,t) \cdotp \log{(\rho(t,\gamma_x,j))}
 \end{equation} 

In this equation, we iterate through each document $j$ that is associated with $t$ and consider the contribution of $j$ to $t$ and also the co-occurrences of token $t$ and center $\gamma_x$ through document $j$. We utilize the log function to constrain the impact of the co-occurrence factor.

\section{Security Analysis}\label{sec: seca}
\name~is applicable in searchable encryption systems and, more generally, in encryption-based document retrieval systems. We envision that the systems utilize \name~are composed of at least two tiers (\eg client and cloud tiers in Figure \ref{sd}). Only the actions performed on the client tier are considered trustworthy and clouds are considered honest but curious. We consider internal and external attackers desire to unveil the encrypted tokens and documents. In addition, attacks are possible in the communication between the client tier and cloud tier. To explain the threat model, we provide the following preliminaries:

\textit{View}: The term `view' denotes the portion that is visible to the cloud during any given interaction between client and server. The index and the set of clusters, the encrypted query (trapdoor) of a given search query ($Q'$), and the collection of encrypted documents $D'$. In some models, $Q'$ also contains a particular weight for each term. The search result set for $Q'$ is considered as $I_c$. Let $V(I_c)$ be this view.

\textit{Trace}: It denotes the data exposed about $I_c$. Our aim is to minimize the data an attacker can infer from $I_c$.  

The View and Trace enclose all the information that the attacker would gain. To encrypt the document set we use probabilistic encryption model that is considered to be one of the most secure encryption techniques \cite{S3BD}. The probabilistic encryption does not utilize one-to-one mapping and so, $D'$ is not prone to dictionary-based attacks \cite{dictionaryAttacks}. In addition, each token of the cluster is deterministically encrypted. Each cluster in the View only shows frequency of the encrypted tokens in documents in plain-text format. We note that even the frequency information can be stored in encrypted form using homomorphic encryption techniques~\cite{homomorphic:slow}, however, due to practical issues and maintain the real-time behavior of the system, we consider that as a future study.

If an attacker gains the access to the system, only he could understand the importance of a particular encrypted token by observing its frequency, but he/she cannot decrypt the token. 

In an extreme scenario, let us assume that an attacker manages access to the key for deterministic encryption from the user's side. Theoretically, the attacker could build a dictionary considering all tokens from the clusters. Eventually, he/she tries to build a clone document set $D''$ utilizing the dictionary. Although all of the tokens extracted from a particular document are sufficient to understand the topic of the document, it is not possible to unveil the whole document. 

\section{Performance Evaluation}
\label{sec:evltn}
\subsection{Experimental Setup and Datasets}
We implemented \name~within the S3BD context to cluster tokens in its central index. For evaluation,  we compare and analyze the quality of clustering against other approaches that are both in encrypted and unencrypted domains. Our implementation of \name~is publicly available in \textcolor{blue}{\url{https://git.io/fjf5X}}. 
For experiment, we used two 10-core 2.8 GHz E5 Xeon processors with 64 GB memory. 
To assure effectiveness of \name, we evaluate it using three distinct datasets. These are selected based on their volume and characteristics of their data. To evaluate performance of \name~with big data, we use a subset of the \texttt{Amazon Common Crawl Corpus (ACCC)} dataset\cite{comcrawl}. The whole dataset size $\approx$150 terabytes. The dataset is not domain-specific and contains different web contents, such as blogs and social media contents. Due to the limited processing capability of our available machine, we sampled from the dataset and randomly selected 6,119 documents that form a $\approx$500 GB document set. The second dataset, named \texttt{Request For Comments (RFC)}, is domain-specific and includes documents about Internet and communication networks. \texttt{RFC} includes 2,000 documents and the size is $\approx$ 247 MB. The third dataset is called \texttt{BBC} that is not domain-specific and includes popular news in various fields such as technology, politics, sports, and business. It contains 1000 documents and is $\approx$ 5 MB. The reason for choosing this small dataset is that, unlike \texttt{ACCC} and \texttt{RFC}, each document of \texttt{BBC} is short and we can verify clusters' coherency manually.
For each dataset, the documents are passed through \textit{Maui} keyword extractor \cite{Maui} to identify keywords semantically represent the document. 


To evaluate \name, we instrument the pre-trained \textit{Google News Word2vec} \cite{mikolov2013efficient} model that determines the similarity between any two words. The model is a 300-dimension vector representation of three million words and phrases. 
Word2vec system requires a text dataset as input to build a vocabulary from the input dataset and learns vector representation of words. The model uses cosine similarity and provides the similarity score ($-1\leq similarity\; score\leq 1$) for any two given tokens. We note that Word2vec model operates only on unencrypted tokens. Hence, to utilize Word2vec, we had to cluster the unencrypted tokens. That means, for evaluation purposes, we build the central index using plain-text tokens but \name~assumes tokens to be encrypted and does not use the plain-text form. 

The performance metric that we use to evaluate the quality of topic-based clustering is overall \emph{cluster coherency}. It represents how tokens in a cluster are related to a certain topic. For a given cluster, \emph{coherence} is calculated based on mean of semantic similarity across all possible pairs of tokens in that cluster. Then, the mean of coherency across all clusters is defined as the overall cluster coherency.

\subsection{Experimental Results}

\subsubsection{\textbf{Clusters' Coherency}}
In this experiment, we determine the coherency of clusters resulted from \name~and compare it against those obtained from three other approaches. One is the conventional $k$-means (termed \emph{baseline}). The second is an enhanced $k$-means, known as \textit{Wordnet}~\cite{encry1}, that generates a synonym set (termed Wordnet synsets) based on the input documents \cite{millerwordnet}. 
Then, $k$-means clustering is performed on the set of synsets. Token distribution in Wordnet is performed using edge counting method proposed by Wu \& Palmer~\cite{millerwordnet}. Both of these approaches operate on unencrypted data. The third approach is \textit{Progeny}~\cite{hu2015progeny}, which is a cluster estimation approach. To make Progeny comparable with others, we input its estimated $k$ to Algorithm~\ref{alg:centr} to select clusters' centers.
 \begin{figure} 
	\centering
	\includegraphics[width=.98\linewidth]{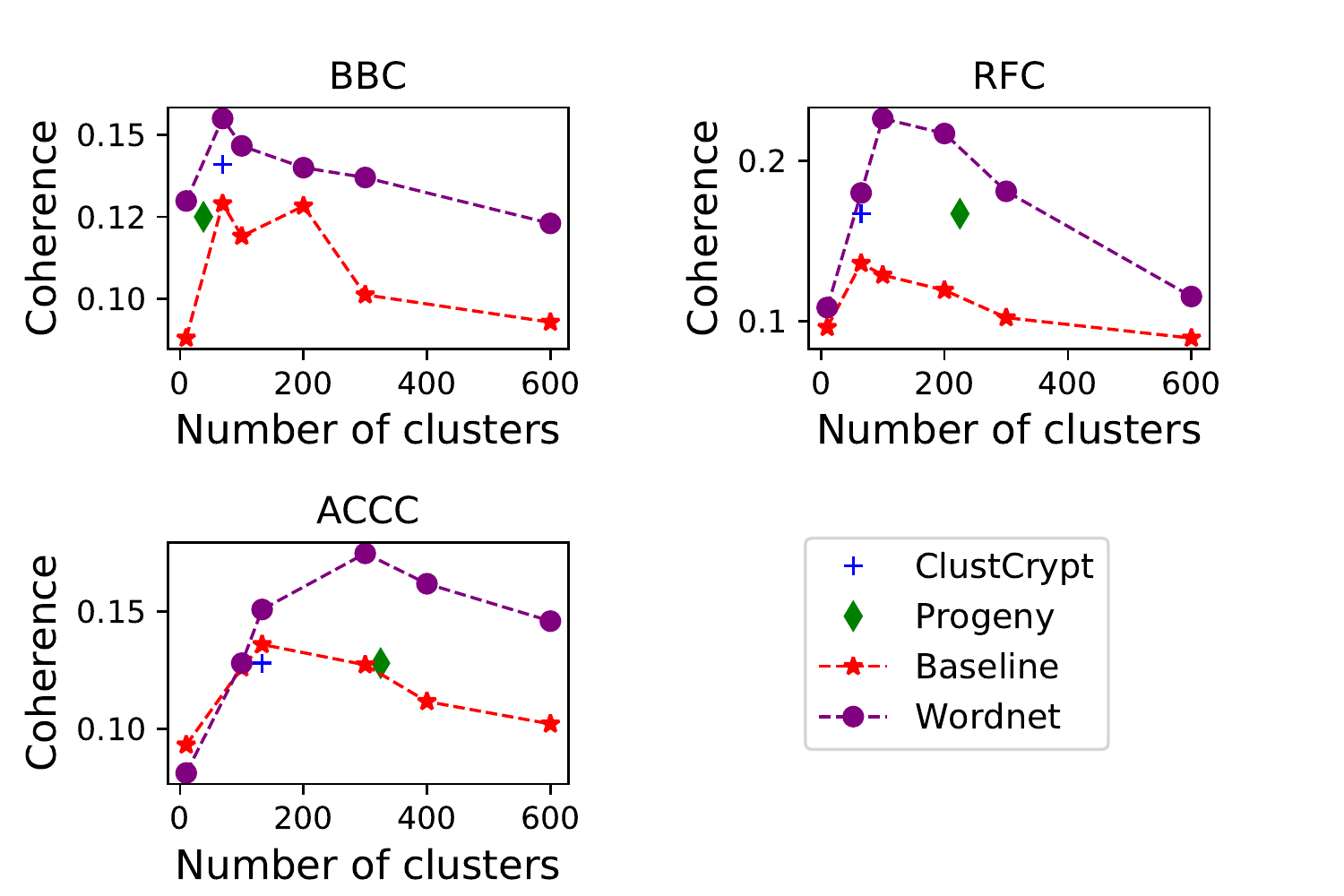}
	\caption{\small{Cluster coherency of \name~against encrypted (Progeny) and unencrypted (Baseline and Wordnet) approaches}}
	\label{fig: pro_unen1}
\end{figure}



Figure~\ref{fig: pro_unen1} shows results of the evaluation on the datasets. We note that both in baseline and Wordnet, $k$ values are randomly chosen. 
We calculate and present clusters' coherency for all considered $k$ values. Thus, for these two approaches, we have multiple data points in the figure. In contrast, \name~and Progeny are not iterative and have one data point. 
Using \name, we obtain 133, 65, and 69 as $k$ values for \texttt{BBC}, \texttt{RFC}, and \texttt{ACCC}, respectively. As \texttt{ACCC} is the largest 
and not domain specific, it yields the highest $k$. On the contrary, although \texttt{RFC} is not the smallest dataset, because it is domain specific, it yields the lowest $k$. We observe that \name~results into the highest coherency for the \texttt{RFC} dataset. 
\textit{Progeny} estimates 42, 225, and 340 clusters for \texttt{BBC}, \texttt{RFC}, and \texttt{ACCC}, respectively. However, after using Algorithm~\ref{alg:centr} for center selection and clustering, only 42, 65, and 69 clusters are built. 
According to the figure, Wordnet clusters have the highest coherency metric. In fact, it is difficult for an encrypted clustering approach (\name) to outperform the unencrypted ones, because they do not have access to the semantic dictionary~\cite{millerwordnet}. In contrast, in \name, as the original meaning of data is unavailable, we populate clusters based on the co-occurrences. We observe that \name~competes with the baseline approach. In particular, in compare to the baseline approach, \name~provides a higher coherency for \texttt{RFC} and \texttt{BBC} datasets. 

\subsubsection{\textbf{Analyzing the Impact of Dynamic Clustering}}
One goal of this research is to enhance performance of S3BD secure search system. As such, we implemented \name~within the context of S3BD and compared the coherency of resulting clusters with its original clustering approach, which is a pre-determined value for $k=10$. Also, the center selection is only based on co-occurrences. In this experiment, we intend to evaluate the improvement that \name~achieves, when it is used within S3BD on the three aforementioned datasets. In this experiment, $k$ estimated for \texttt{BBC, RFC} and \texttt{ACCC} are 69, 65, and 133, respectively.

\begin{figure}[H]
	\centering
	\includegraphics[width=.9\linewidth]{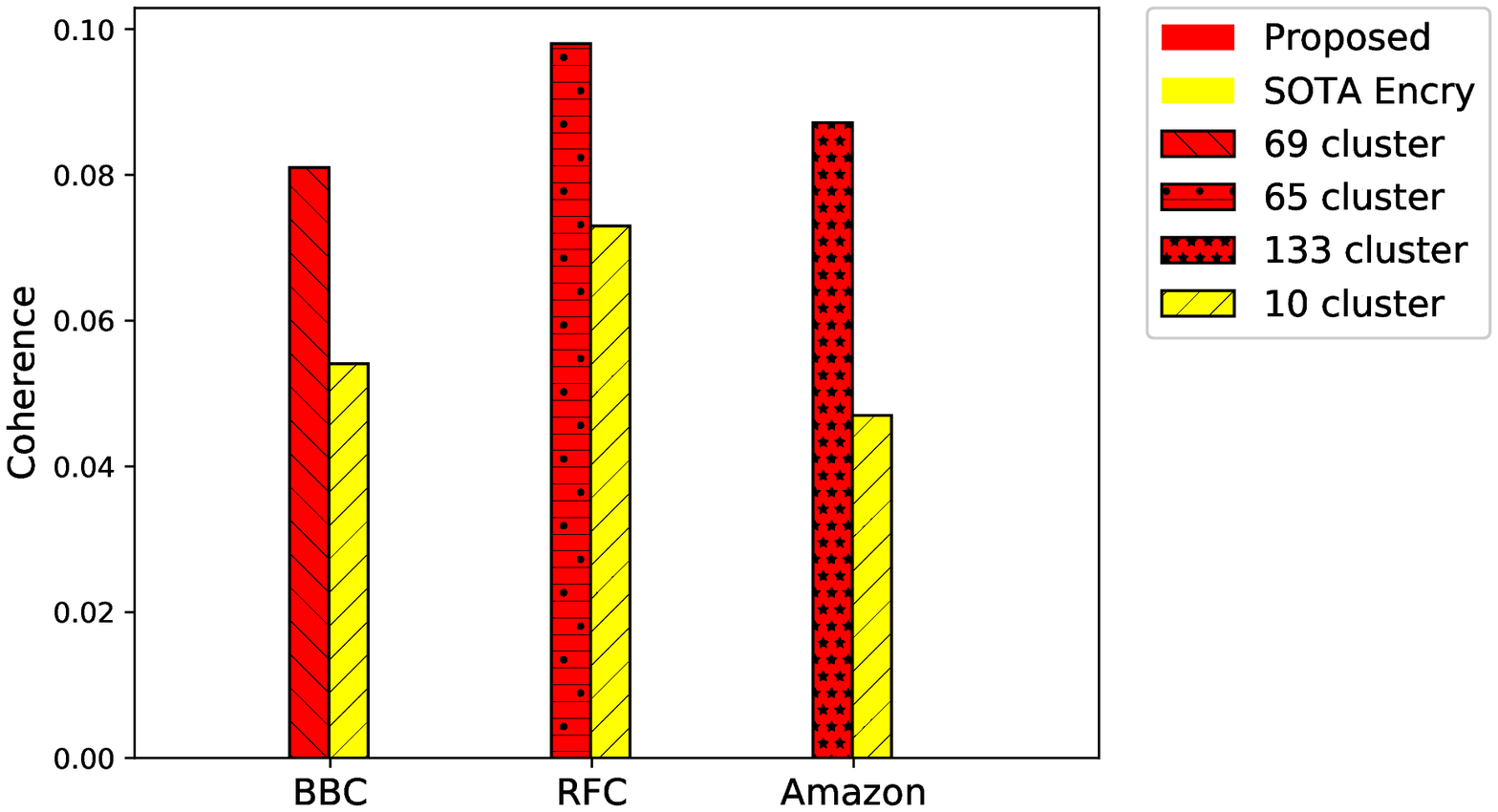}
	\caption{\small{Comparing the impact of dynamic clustering of \name~with static clustering in S3BD secure search system}}
	\label{fig: Pro_en}
\end{figure}

Figure \ref{fig: Pro_en} shows that for all the studied datasets, clusters generated by \name~have remarkably higher coherency than the original static approach. This shows determining number of cluster dynamically based on dataset characteristics and choosing center tokens based on the centrality concept is effective. Such efficiency leads to more accurate and relevant semantic search results on big data, because tokens in clusters are more related to the topic of the cluster. For further analysis, next experiments concentrate on the impact of \name~on the quality of search results.

\subsubsection{\textbf{Analyzing the Impact of \name~on Search Systems}}
Quality of clustering tokens impacts both the accuracy and response time (\ie search time) of the search systems like S3BD. In fact, the aim of this study is to improve the clusters' coherency that impacts the search accuracy by retrieving more relevant documents. To evaluate the clustering impact, in this part, we compare and analyze how the search accuracy of S3BD system is affected by utilizing \name's clusters against the circumstance where the original static clustering is utilized. Then, we study the impact of clustering on the search time too. For evaluation, we generated a set of 10 benchmark search queries for each dataset, as listed in Table~\ref{bench}. 

\begin{table}[H]

\centering

\begin{tabular}{|p{2.2cm}|p{2.89cm}|p{2.4cm}|}

\hline

\scriptsize{\texttt{\textbf{ACCC}}  }                  & \scriptsize{\texttt{\textbf{BBC}}}                & \scriptsize{\texttt{\textbf{RFC}}}                   \\ \hline

\texttt{\scriptsize{Orlando Magic}}                  & \texttt{\scriptsize{News Update}}        & \texttt{\scriptsize{Internet}}              \\ \hline

\texttt{\scriptsize{Samsung Galaxy}}         & \texttt{\scriptsize{Top Movies}}             & \texttt{\scriptsize{TCP}}                   \\ \hline

\texttt{\scriptsize{Baseball routine}}        & \texttt{\scriptsize{Recent Attacks}}        & \texttt{\scriptsize{Fiber Doctor}} \\ \hline

\texttt{\scriptsize{Recommendation}}                & \texttt{\scriptsize{Endangered Animals}} & \texttt{\scriptsize{Wifi}}                  \\ \hline

\texttt{\scriptsize{North America}}     & \texttt{\scriptsize{Score Updates}}             & \texttt{\scriptsize{IoT}}    \\ \hline

\texttt{\scriptsize{Tennis Tournament}} & \texttt{\scriptsize{Champions League}}       & \texttt{\scriptsize{Radio Frequency}}       \\ \hline

\texttt{\scriptsize{Holy Martyr}}       & \texttt{\scriptsize{World Health Issue}}     & \texttt{\scriptsize{UDP}}                   \\ \hline

\texttt{\scriptsize{Library}}            & \texttt{\scriptsize{People and Business }}         & \texttt{\scriptsize{Edge Computing}}        \\ \hline

\texttt{\scriptsize{Stardock}}               & \texttt{\scriptsize{China Market}}          & \texttt{\scriptsize{Encryption schemes}} \\ \hline

\texttt{\scriptsize{Orthodox Church}}    & \texttt{\scriptsize{European Stock Exchange}} & \texttt{\scriptsize{Broadcasting}}          \\ \hline

\end{tabular}

\caption{\small{Benchmark search queries for evaluated datasets}}

\label{bench}
\end{table}

\paragraph*{Impact on relevancy of the search results} 

To measure the relevancy of search results for each query, we use \textit{TREC-Style Average Precision} scoring method~\cite{mariappan}. This method works based on the recall-precision concept and score is calculated by $\sum_{i=0}^N r_i/N$, where $r_i$ denotes the score for $i$th retrieved document and $N$ is the cutoff number (number of search results) that we consider as 10. Therefore, we call it \emph{TSAP@10}. 



\begin{figure}
	\vspace{-10pt}
	\centering
	\includegraphics[width=.7\linewidth]{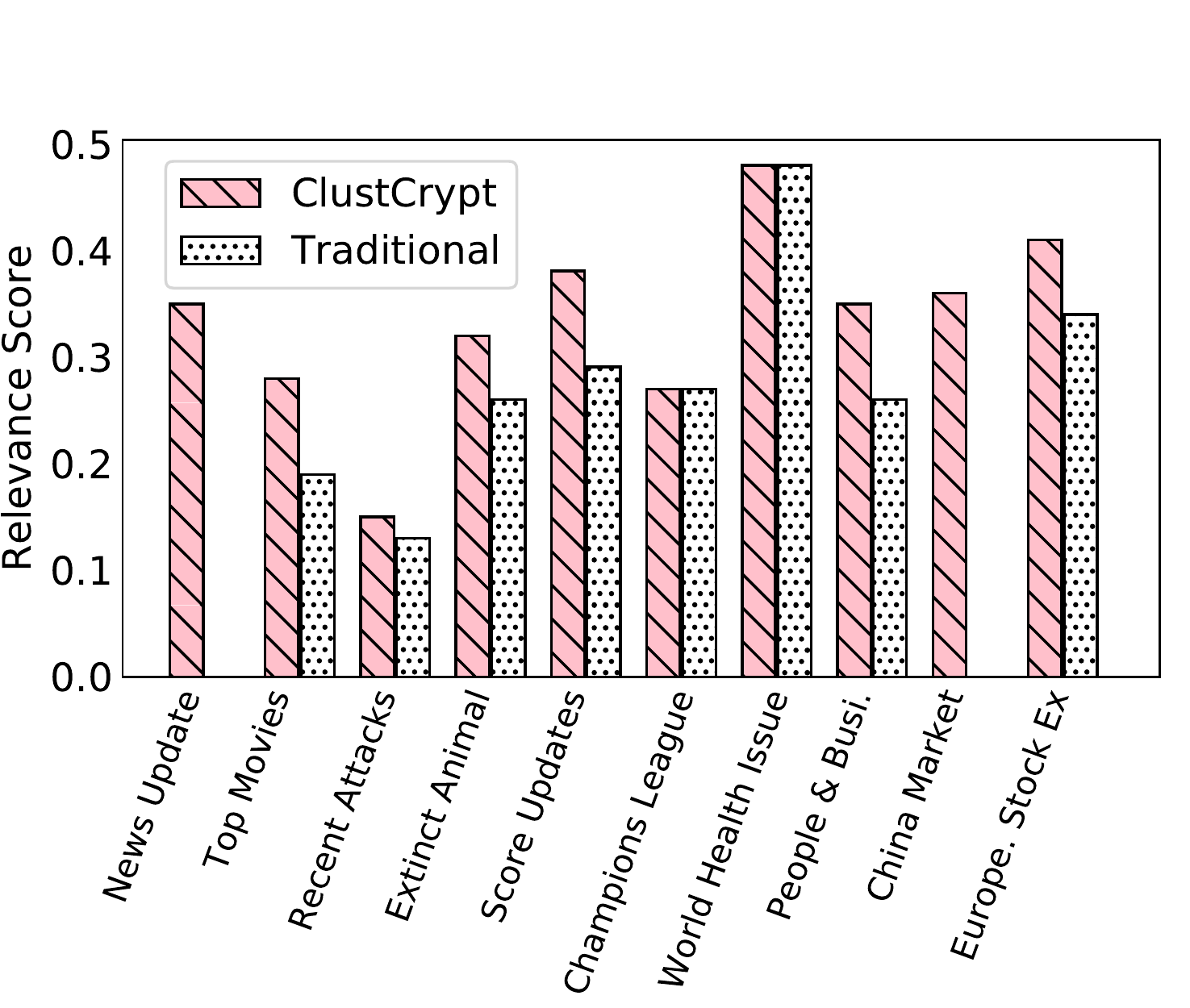}
\vspace{-5pt}
	\caption{\small{Search relevancy comparison based on TSAP@10 scoring metric: \name~vs. original S3BD clustering }}
	\label{fig: search_ac}

\end{figure}

We measure TSAP@10 score only for the \texttt{RFC} dataset and its benchmark queries. The reason is that it is domain-specific and feasible to determine the relevancy of the retrieved documents. 
To compare the relevancy provided by \name~against original S3BD clustering, we search the benchmark queries using S3BD. In Figure~\ref{fig: search_ac}, the relevancy scores (vertical axis) of each query (horizontal axis) by utilizing the two approaches are represented. 
According to the Figure, for most of the queries, the TSAP@10 scores obtained by \name~clusters offer the higher relevancy. For two queries, the same \emph{TSAP@10} score is offered, because the retrieved document lists are equivalent for them. Also, \name~clusters provide score for \textit{news update} and \textit{China Market} benchmark queries, whereas original S3BD clusters do not retrieve any relevant documents for them.

\paragraph*{Impact on the search time}

\begin{figure}[H]
	\centering
	\includegraphics[width=.85\linewidth]{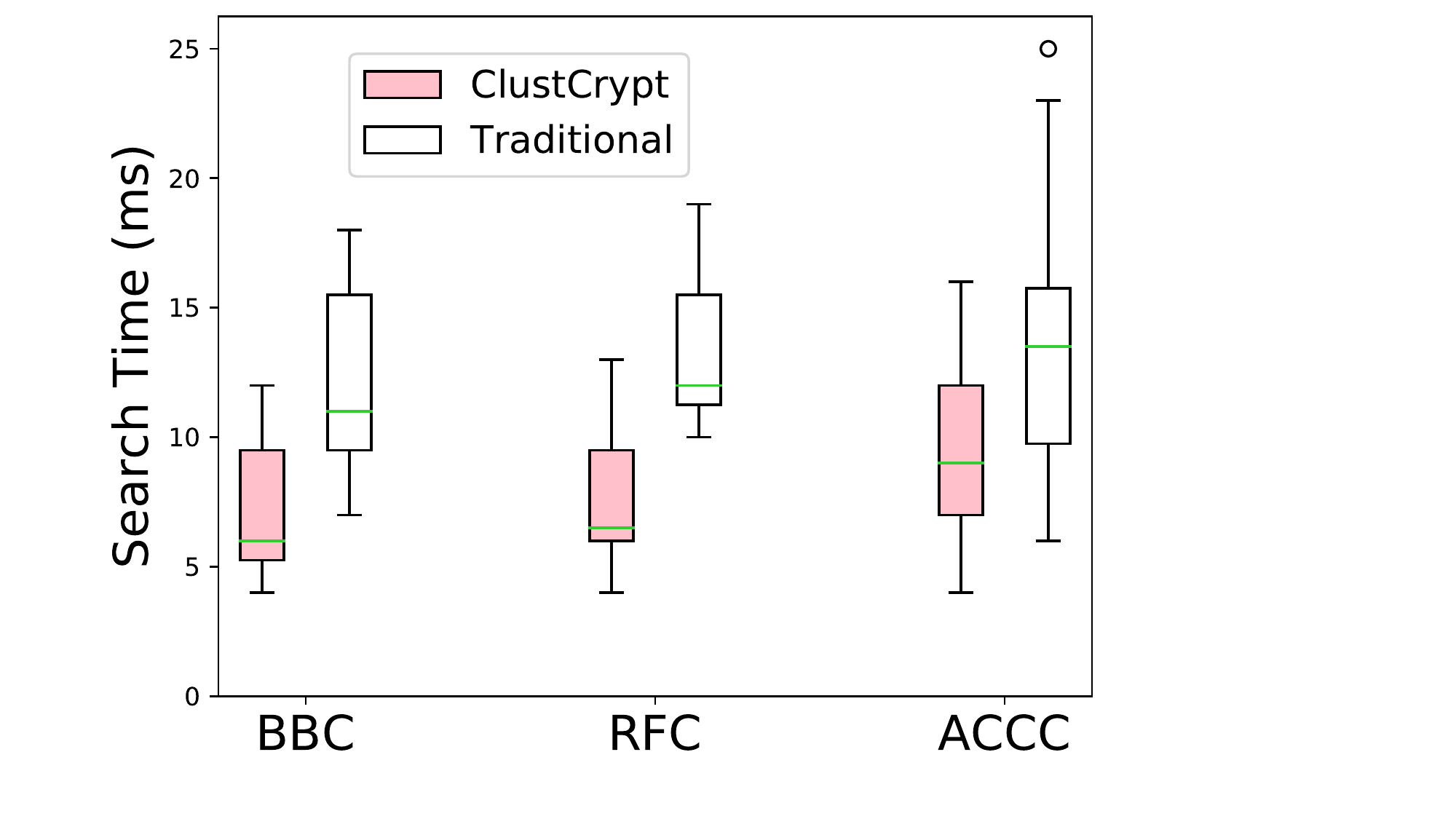}
	\caption{\small{Search time of \name~versus original S3BD clustering}}
	\label{fig: time_en}
\end{figure}

Figure~\ref{fig: time_en} presents the overall search time of the benchmark queries for each dataset. This figure indicates that, in addition to providing more accurate results, \name~also offers a shorter search time. Longer search time impacts scalability and real-time quality of the search operation on big data. However, we can observe that in compare to the clusters generated by original S3BD approach, \name~is more scalable in terms of real-time search operation.

\section{CONCLUSIONS and Future Works}\label{sec:conclsn}
In this paper, we developed \name, to efficiently perform topic-based clustering of unstructured encrypted big data in the cloud.
\name~approximates the number of clusters for a given dataset with reduced time complexity, compared to conventional clustering approaches.
Utilizing statistical meta-data, we obtain the probabilistic tendency of each token being segregated from others and use it to estimate the number of clusters. We leveraged the probabilistic analysis to determine center tokens and disseminate encrypted tokens to relevant clusters. 
Experimental results show that clustering using \name~provides on average 60\% more coherency, comparing to conventional approaches used for encrypted data. We concluded that it is difficult for encrypted clustering to outperform plain-text clustering. However, \name~performs closer to the benchmark than other encryption-based clustering approaches. From the searchable encryption perspective, \name~helps to provide more relevant search results. In future, we plan to extend \name~to cluster growing datasets (\eg social networks). Also, we will investigate cluster pruning mechanisms to improve the search accuracy.

\section*{Acknowledgments}
\small{
We thank reviewers of the manuscript. 
This research was supported by the Louisiana Board of Regents under grant number LEQSF(2017-20)-RD-B-06, National Science Foundation under grant number CNS-1429526, and Perceptive Intelligence, LLC.} 

\linespread{.82}
\bibliographystyle{plain}

\bibliography{references}

\begin{thebibliography}{10}

\bibitem{kcomplexity}
\url{www.researchgate.net/post/What_is_the_time_complexity_of_clustering_algorithms}.
\newblock {Accessed Aug. '18}.

\bibitem{yahooacci}
\url{www.money.cnn.com/2017/10/03/technology/business/yahoo-breach-3-billion-accounts/index.html}.
\newblock Accessed Feb '19.

\bibitem{verizonacci}
\url{www.upguard.com/breaches/verizon-cloud-leak}.
\newblock Accessed Feb '19.

\bibitem{comcrawl}
\url{http://commoncrawl.org}.
\newblock Accessed Feb'18.

\bibitem{encry1}
\url{www.github.com/darenr/wordnet-clusters}.
\newblock {Accessed May. 18}.

\bibitem{cloudacci2}
\url{www.csoonline.com/article/2130877/data-breach/}.
\newblock Accessed Nov. '18.

\bibitem{infoot}
\url{www.entrepreneur.com/article/273561}.
\newblock {Accessed Nov. 18}.

\bibitem{aggarwal2001surprising}
Charu~C Aggarwal, Alexander Hinneburg, and Daniel~A Keim.
\newblock On the surprising behavior of distance metrics in high dimensional
  space.
\newblock In {\em Proceedings of the 8th International conference on database
  theory (ICDT)}, pages 420--434, Jan 2001.

\bibitem{aggarwal2019performance}
Swati Aggarwal, Nitika Agarwal, and Monal Jain.
\newblock Performance analysis of uncertain k-means clustering algorithm using
  different distance metrics.
\newblock In {\em Computational Intelligence: Theories, Applications and Future
  Directions-Volume I}, pages 237--245. 2019.

\bibitem{kullback1951information}
Hirotogu Akaike.
\newblock {\em Journal of Selected Papers of Hirotugu Akaike}, page 199, 2012.

\bibitem{usercentri2019}
Sahan and Ahmad, SM~Zobaed, Raju Gottumukkala, and MA~Salehi.
\newblock Edge computing for user-centric secure search on cloud-based
  encrypted big data.
\newblock In {\em Proceedings of the 21st International Conference on High
  Performance Computing and Communications, {HPCC}}. IEEE, 2019.

\bibitem{Google}
L.~A. Barroso, J.~Dean, and U.~Holzle.
\newblock Web search for a planet: The google cluster architecture.
\newblock {\em IEEE Micro}, 23(2):22--28, Mar. 2003.

\bibitem{kmeans}
Pavel Berkhin.
\newblock A survey of clustering data mining techniques.
\newblock In {\em Grouping multidimensional data}, pages 25--71. 2006.

\bibitem{Can1990}
Fazli Can and Esen~A. Ozkarahan.
\newblock Concepts and effectiveness of the cover-coefficient-based clustering
  methodology for text databases.
\newblock {\em Journal of ACM Trans. Database Syst.}, 15(4):483--517, December
  1990.

\bibitem{cao2014privacy}
Ning Cao, Cong Wang, Ming Li, Kui Ren, and Wenjing Lou.
\newblock Privacy-preserving multi-keyword ranked search over encrypted cloud
  data.
\newblock {\em IEEE Transactions on parallel and distributed systems (TPDPS)},
  25(1):222--233, 2014.

\bibitem{coates2012learning}
A.~Coates and Andrew Ng.
\newblock Learning feature representations with k-means.
\newblock In {\em Neural networks: Tricks of the trade}, pages 561--580. 2012.

\bibitem{cutting2017scatter}
Douglass~R Cutting, David~R Karger, Jan~O Pedersen, and John~W Tukey.
\newblock Scatter/gather: A cluster-based approach to browsing large document
  collections.
\newblock 51(2):148--159, 2017.

\bibitem{homomorphic:slow}
L{\'e}o Ducas and Daniele Micciancio.
\newblock In {\em Proceedings of 34th Annual International Conference on the
  Theory and Applications of Cryptographic Techniques}, pages 617--640, 2015.

\bibitem{comjnl/32.3.193}
P.~Hammersley.
\newblock Editorial – information and information systems.
\newblock {\em The Computer Journal}, 32(3):193, 1989.

\bibitem{hu2015progeny}
Chenyue~W Hu, Steven~M Kornblau, John~H Slater, and Amina~A Qutub.
\newblock Progeny clustering: a method to identify biological phenotypes.
\newblock {\em Journal of Scientific reports}, 5:12894, 2015.

\bibitem{S3BD}
M.~A.~Salehi J.~Woodworth.
\newblock {S3BD}: Secure semantic search over encrypted big data in the cloud.
\newblock {\em Journal of Concurrency and Computation:Practice and Experience
  ({CCPE})}, 28, 2018.

\bibitem{lee2009pca}
Chih Lee, Ali Abdool, and Chun-Hsi Huang.
\newblock Pca-based population structure inference with generic clustering
  algorithms.
\newblock {\em Journal of BMC bioinformatics}, 10(1):S73, Jan. 2009.

\bibitem{LiuCroft}
X.~Liu and W.~Bruce Croft.
\newblock Cluster-based retrieval using language models.
\newblock In {\em Proceedings of the 27th International ACM SIGIR Conference on
  Research and Development in Information Retrieval}, SIGIR '04, 2004.

\bibitem{lleti2004selecting}
R~Llet{\i}, M~Cruz Ortiz, Luis~A Sarabia, and M~Sagrario S{\'a}nchez.
\newblock Selecting variables for k-means cluster analysis by using a genetic
  algorithm that optimises the silhouettes.
\newblock {\em Journal of Analytica Chimica Acta}, 515(1):87--100, 2004.

\bibitem{mariappan}
A.~K. Mariappan, R.~M. Suresh, and V.~Subbiah Bharathi.
\newblock A comparative study on the effectiveness of semantic search engine
  over keyword search engine using tsap measure.
\newblock {\em Journal of Computer Applications EGovernance and Cloud Computing
  Services}, pages 4--6, Dec. 2012.

\bibitem{Maui}
Olena Medelyan, Eibe Frank, and Ian~H. Witten.
\newblock Human-competitive tagging using automatic keyphrase extraction.
\newblock In {\em Proceedings of the 14th Conference on Empirical Methods in
  Natural Language}, EMNLP '09, pages 1318--1327, Aug. 2009.

\bibitem{mikolov2013efficient}
Tomas Mikolov, Ilya Sutskever, Kai Chen, Greg~S Corrado, and Jeff Dean.
\newblock Distributed representations of words and phrases and their
  compositionality.
\newblock In {\em Proceedings of the 27th Conference on Neural Information
  Processing Systems, {(NIPS)}}, pages 3111--3119, Dec 2013.

\bibitem{millerwordnet}
George~A Miller.
\newblock Wordnet: a lexical database for english.
\newblock {\em Journal of Communications of the ACM}, 38(11):39--41, 1995.

\bibitem{Paparrizos}
John Paparrizos and Luis Gravano.
\newblock k-shape: Efficient and accurate clustering of time series.
\newblock In {\em Proceedings of the 2015 ACM SIGMOD International Conference
  on Management of Data}, SIGMOD '15, 2015.

\bibitem{pelleg2000x}
Dan Pelleg and Andrew~W. Moore.
\newblock X-means: Extending k-means with efficient estimation of the number of
  clusters.
\newblock In {\em Proceedings of the 17th International Conference on Machine
  Learning}, ICML '00, pages 727--734, 2000.

\bibitem{pham2018survey}
H.~Pham, J.~Woodworth, and M.~A. Salehi.
\newblock Survey on secure search over encrypted data on the cloud.
\newblock {\em Accepted in Journal of Concurrency and Computation:Practice and
  Experience (CCPE)}.

\bibitem{salehi2017reseed}
Mohsen~Amini Salehi, Thomas Caldwell, Alejandro Fernandez, Emmanuel Mickiewicz,
  Eric~WD Rozier, Saman Zonouz, and David Redberg.
\newblock Reseed: A secure regular-expression search tool for storage clouds.
\newblock {\em Journal of Software: Practice and Experience}, 47(9):1221--1241,
  2017.

\bibitem{vrahatis2002new}
Michael~N Vrahatis, Basilis Boutsinas, Panagiotis Alevizos, and Georgios
  Pavlides.
\newblock The new k-windows algorithm for improving thek-means clustering
  algorithm.
\newblock {\em journal of complexity}, 18(1):375--391, 2002.

\bibitem{dictionaryAttacks}
Ding Wang and Ping Wang.
\newblock Offline dictionary attack on password authentication schemes using
  smart cards.
\newblock In {\em Information Security}, 2015.

\bibitem{S3C}
J.~Woodworth, M.~A. Salehi, and V.~Raghavan.
\newblock {S3C}: An architecture for space-efficient semantic search over
  encrypted data in the cloud.
\newblock In {\em Proceedings of the 4th IEEE International Conference on Big
  Data}, Big Data'16, pages 3722--3731, 2016.

\bibitem{XuCroft}
Jinxi Xu and W.~Bruce Croft.
\newblock Cluster-based language models for distributed retrieval.
\newblock In {\em Proceedings of the 22nd International ACM Conference on
  Research and Development in Information Retrieval}, SIGIR '99, pages
  254--261, Aug. 1999.

\bibitem{nocs2}
SM~Zobaed, Md~Enamul Haque, Shahidullah Kaiser, and Razin~Farhan Hussain.
\newblock Nocs2: Topic-based clustering of big data text corpus in the cloud.
\newblock In {\em Proceedings of the 21st International Conference of Computer
  and Information Technology (ICCIT)}, pages 1--6, Dec. 2018.

\bibitem{zobaedbig}
Sm~Zobaed and Mohsen~Amini Salehi.
\newblock Big data in the cloud.
\newblock In {\em Encyclopedia of Big Data {. Edited by Albert Zomaya and
  Sherif Sakr, Springer International Publishing}}.

\end{thebibliography}


\balance

\end{document}